\documentclass[12pt]{article}
\usepackage{graphicx}
\renewcommand{\theequation}{\arabic{equation}}
\renewcommand{\thesection}{\arabic{section}}
\renewcommand{\thefootnote}{\fnsymbol{footnote}}

\newcommand{\vs}[1]{\vspace{#1 mm}}

\renewcommand{\a}{\alpha}
\renewcommand{\b}{\beta}

\newcommand{\e}{\epsilon}

\begin{document}
\noindent
\topmargin 0pt
\oddsidemargin 5mm

\begin{titlepage}
\setcounter{page}{0}
\thispagestyle{empty}
\begin{flushright}
July, 2004\\
OU-HET 468\\
hep-ph/0402106\\
\end{flushright}
\vs{4}
\begin{center}
{\LARGE{\bf The role of Majorana CP phases in 
the bi-maximal mixing scheme
-hierarchical Dirac mass case-}} \\ 
\vs{6}
{\large 
Tetsuo Shindou\footnote{
Present address: Theory Group, KEK, Tsukuba, Ibaraki 305-0801, Japan;
e-mail address: shindou@post.kek.jp} and 
Eiichi Takasugi\footnote{e-mail address:
takasugi@het.phys.sci.osaka-u.ac.jp}\\
\vs{2}
{\em Department of Physics,
Osaka University \\ Toyonaka, Osaka 560-0043, Japan} \\
}
\end{center}
\vs{6}
\centerline{{\bf Abstract}}
We discuss the energy scale profile of the bi-maximal mixing 
which is given at the GUT energy scale in 
the minimal SUSY model, associated with an assumption that 
$Y_\nu^\dagger Y_\nu$ is diagonal, where $Y_\nu$ is the 
neutrino-Yukawa coupling matrix. In this model, the Dirac 
mass matrix which appears in the seesaw neutrino mass matrix 
is determined by three neutrino masses, two relative Majorana 
phases and three heavy Majorana masses. All CP phases are 
related by two Majorana phases. We show that the requirement that 
the solar mixing angle moves from the maximal mixing at GUT to 
the observed one as the energy scale decreases by 
the renormalization effect. We discuss the leptogenesis, and 
the lepton flavor violation process by assuming 
the universal soft breaking terms.

\end{titlepage}

\newpage
\renewcommand{\thefootnote}{\arabic{footnote}}
\setcounter{footnote}{0}

\section{Introduction} 
The bi-maximal mixing scheme[1] may be most attractive one. 
It has a simple and beautiful 
structure and there are various models which give the bi-maximal mixing
at the GUT scale[2].
In addition 
to this, we feel the property $V_{13}=0$ is interesting. 
If the bi-maximal 
mixing is realized at the GUT scale $M_X$, the Dirac CP phase $\delta$ as 
well as $|V_{13}|$ which are absent at $M_X$ are induced following to the 
renormalization group equation at the low energy. This may give us a chance 
to predict these quantities. 
Another interesting point is that 
we may able to solve the discrepancy 
between the maximal solar mixing angle at the GUT scale, 
$\tan^2 \theta_\odot=1$ and the experimental data[3,4] 
at the low energy
\begin{eqnarray}
\tan^2 \theta_\odot\simeq 0.40\;.
\end{eqnarray}

Let us consider the renormalization group equation due to 
the neutrino-Yukawa and the $\tau$-Yukawa couplings.  We 
have shown[5,6] that the effect due to the $\tau$-Yukawa coupling rotates 
the solar angle toward the dark side. Therefore, the large 
neutrino-Yukawa couplings are needed[6,7] to compensate this 
and rotates the solar angle toward the normal side. 
In this analysis, Majorana CP phases[8] 
in the neutrino mixing matrix[9] play 
an important role. 

In this paper, we continue this analysis further by considering the 
neutrino mass matrix derived through the seesaw mechanism in the 
framework of the MSSM with the universal soft supersymmetry breaking terms. 
Our main motivation is to examine the the structure of the Dirac mass matrix 
and explore the possible relation of CP phases at the high and 
low energies. In general, the Dirac mass matrix introduces new 
CP phases and there is no relation between CP phases in the 
low energy and the high energy[10].

In Sec.2, we explain the assumptions which we adopt to construct 
the neutrino mass matrix. The renormalization 
group analysis is briefly explained and a typical form of the neutrino-Yukawa 
couplings is discussed. By assuming this typical from, 
the Dirac mass is determined. In Sec.3, 
the various results including the asymmetry parameter of the leptogenesis, 
the lepton flavor violation are discussed.  The numerical analysis is 
presented in Sec.3.  In Sec.4, the summary and discussion are 
given.

\section{The model}

We assume the MSSM with the neutrino-Yukawa coupling matrix, $Y_\nu$  
and the right-handed Majorana neutrino mass matrix, $M$.
The related terms are given by
\begin{eqnarray}
{\cal{L}}_{y+M}={\overline{N_R}}\phi_u^\dagger Y_\nu\nu_L -
\frac12 {\overline{(N_R)^C}}M N_R+h.c.\;.
\end{eqnarray}
The soft SUSY breaking terms are assumed to be 
universal and the source of the lepton flavor violation (LFV) 
is only through the neutrino-Yukawa couplings.  
The left-handed neutrino mass matrix, $m_\nu$ is derived through 
the seesaw mechanism as 
\begin{eqnarray}
m_\nu(M_X)=m_D^T M^{-1} m_D\;, 
\end{eqnarray}
where 
\begin{eqnarray}
m_D=Y_\nu \frac{v_u}{\sqrt 2}
\end{eqnarray}
with $v_u=v\sin \beta$. Here, the neutrino mass matrix is effectively 
given at 
the right-handed neutrino mass scale, while  
$Y_\nu$ and $M$ in Eq.(3) are defined at the GUT scale, $M_X$.  

The Dirac mass matrix is generally expressed by
\begin{eqnarray}
m_D=V_R^\dagger D_D V_{L}\;,
\end{eqnarray}
where $V_R$ and $V_L$ are unitary matrices, 
$D_D$ is a diagonal mass matrix
\begin{eqnarray}
D_D={\rm diag}(m_{D1},m_{D2},m_{D3})\;,
\end{eqnarray}
with real, positive eigenvalues, $m_{Di}$.  
In the following, we take 
the diagonal basis of $M$, 
\begin{eqnarray}
M=D_R={\rm diag}(M_1,M_2,M_3)\;,
\end{eqnarray}
with real positive eigenvalues, $M_i$. 

In this paper, we consider the hierarchical Dirac mass case, 
\begin{eqnarray}
m_{D3}\gg m_{D2}\gg m_{D1}\;. 
\end{eqnarray}
Also we take 
\begin{eqnarray}
M_3>M_2>M_1\;.
\end{eqnarray}

\begin{enumerate}

\item[(1)] The model

Our mode consists of the following contents. 
\begin{enumerate}
	\item[(A.1)] The bi-maximal mixing is realized at the GUT scale, $M_X$.
	\item[(A.2)] The experimental solar mixing angle is achieved 
	by the renormalization group effect due to the neutrino-Yukawa 
	and the $\tau$-Yukawa couplings. 
	\item[(A.3)] $Y_\nu^\dagger Y_\nu$ is assumed to be a diagonal matrix.
\end{enumerate}

From (A.1), the neutrino mass matrix, $m_\nu$ at $M_X$ is given by 
\begin{eqnarray}
m_\nu(M_X)=O_BD_\nu O_B^T \;,
\end{eqnarray}
where $O_B$ is the bi-maximal mixing matrix
\begin{eqnarray}
O_B=\pmatrix{
\frac{1}{\sqrt{2}}&-\frac{1}{\sqrt{2}}&0\cr
\frac{1}{2}&\frac{1}{2}&-\frac{1}{\sqrt{2}}\cr
\frac{1}{2}&\frac{1}{2}&\frac{1}{\sqrt{2}}\cr}\;. 
\end{eqnarray}
$D_\nu$ is a diagonal matrix with complex eigenvalues,
\begin{eqnarray} 
D_\nu={\rm diag}(m_1,m_2,m_3)={\rm diag}
(|m_1|,|m_2|e^{i\a_o},|m_3|e^{i\b_o})\;,
\end{eqnarray}
where $\a_0$ and $\b_0$ are Majorana phases[8].

\item[(2)] The motivation of the assumption (A.3) 
		
In our papers[5,6], we showed that the $\tau$-Yukawa contribution 
rotates the solar angle toward the dark side. Therefore, 
(A.2) requires[6] that the contributions from 
the neutrino-Yukawa should compensate the $\tau$-Yukawa effect 
and rotate the solar angle into the normal side. For this, 
some elements of $Y_\nu^\dagger Y_\nu$ must large. 
On the other hand, 
the LFV processes take place through mixings in the slepton sector[11]. 
The rate for $\ell_i \to 
\ell_j+\gamma$ is proportional to $|(Y_\nu^\dagger Y_\nu)_{ij}|^2$, 
and if we require that 
its branching ratio is less than, say, $10^{-12}$, then 
$|(Y_\nu^\dagger Y_\nu)_{ij}|<3\times 10^{-3}(1/\tan \b)$ ($i\neq j$) 
is required. 
Thus, only the elements which can be large are 
diagonal elements of  $Y_\nu^\dagger Y_\nu$, which leads to 
the the assumption (A.3). The possible modification of this assumption 
will be mentioned later. 

\item[(3)] The form of $m_D=(v_u/\sqrt 2)Y_\nu$

In the following, we sometimes use $m_D$ instead of $Y_\nu$ for 
the convenience. From the assumption (A.3), $m_D$ is 
expressed by
\begin{eqnarray}
m_D=V_R^\dagger D_D P_{ex}\;,
\end{eqnarray}
where $V_R$ is a unitary matrix, 
$D_D$ is a diagonal matrix defined in Eq.(6) 
and $P_{ex}$ is the matrix to exchange the eigenvalues of 
$D_D$. 

The matrix $P_{ex}$ is fixed by considering 
the renormalization group effect to the solar neutrino mixing 
parameters. We take
\begin{eqnarray}
P_{ex}=\pmatrix{0&0&1\cr 0&1&0\cr -1&0&0\cr}\;,
\end{eqnarray}
from the reason we explained later.

\item[(4)] The renormalization group effect

The renormalization group equation is given 
for $M_X>\mu>M_R$ by 
\begin{eqnarray}
\frac{d m_\nu}{d \ln \mu}=\frac{1}{16\pi^2}\left\{ 
[(Y_\nu^\dagger Y_\nu)^T + (Y_e^\dagger Y_e)^T ]m_\nu
+m_\nu [(Y_\nu^\dagger Y_\nu) + (Y_e^\dagger Y_e)]
\right\},
\end{eqnarray}
aside from the terms proportional to the unit matrix. Here, 
$M_R$ is the right-handed neutrino mass scale. 
For the charged lepton 
Yukawa coupling matrix, we consider 
only the $\tau$-Yukawa coupling,
$Y_e={\rm diag}(0,0,y_\tau)$. 
When $\mu<M_R$, 
only the $\tau$-Yukawa couplings contribute, because the heavy 
neutrinos decouple from the interaction.

From the assumption (A.3), we express 
\begin{eqnarray}
Y_\nu^\dagger Y_\nu={\rm diag}(y_1^2,y_2^2,y_3^2)\;. 
\end{eqnarray}
The contribution from it is split into the one proportional 
to the unit 
matrix, say, ${\rm diag}(0,y_2^2,0)$ and the rest, 
${\rm diag}(y_1^2-y_2^2,0,y_3^2-y_2^2)$. The former contributes 
the overall normalization of neutrino masses so that we discard it. 
As a result, the 
renormalization equation is expressed in a good approximation 
as
\begin{eqnarray}
m_{\nu}(m_Z)=m_{\nu}(M_X)+K^T m_{\nu}(M_X)+m_{\nu}(M_X)K\;,
\end{eqnarray}
where[6] 
\begin{eqnarray}
K=\pmatrix{\e_e & 0 & 0\cr 0&0&0 \cr 0&0&\e_\tau \cr}\;.
\end{eqnarray} 
The $\e_e$ and $\e_\tau$ are given explicitly by 
\begin{eqnarray}
\e_e&=&\frac{y_1^2-y_2^2}{16\pi^2}\ln\left(
\frac{m_X}{M_R}\right)\;,\nonumber\\
\e_\tau&=&\frac{y_3^2-y_2^2}{16\pi^2}\ln\left(
\frac{m_X}{M_R}\right)
+\frac{y_\tau^2}{16\pi^2}\ln\left(
\frac{m_X}{M_Z}\right)\;.
\end{eqnarray}
Here we neglect the threshold effect of $M_i$, which we discuss 
later and take $M_i=M_R$.

Since the renormalization group effect is discussed in detail 
in Refs.4 and 5, 
we give only the result.  
The effect to the sizes of neutrino masses, the atmospheric 
mass squared difference and the atmospheric mixing angle are small 
for $|m_i|\le 0.1$eV. Only the effect appears to the solar mixing 
angle and the solar mass squared masses, which are related by
\begin{eqnarray}
\tan^2\theta_{\odot}=
\frac{1+(\epsilon_{\tau}-2\epsilon_e)
\cos^2(\alpha_0/2)m_1^2/\Delta m_{\odot}^2}
{1-(\epsilon_{\tau}-2\epsilon_e)
\cos^2(\alpha_0/2)m_1^2/\Delta m_{\odot}^2}\;,
\end{eqnarray}
where $\theta_{\odot}$ and $\Delta m_{\odot}^2$ are the experimental 
values of the solar mixing angle and the mass squared difference 
which are defined at the low energy scale, $m_Z$.
In order to obtain $\tan^2\theta_{\odot}\simeq 0.40$, it is 
required
\begin{eqnarray}
(2\epsilon_{e}-\epsilon_\tau)
\cos^2(\alpha_0/2)(m_1^2/\Delta m_{\odot}^2)=\cos 2\theta_\odot \;.
\end{eqnarray}
This equation 
gives the constraint on $y_i^2 m_1^2$. Therefore, the smaller 
neutrino masses requires the larger Yukawa couplings. Since 
the Yukawa couplings can not be very large, the neutrino masses 
must be large. If we take $m_1= 0.05$eV, 
$\Delta m_{\rm sol}^2=6.9\times 10^{-5}{\rm eV}^2$, 
$\cos 2\theta_\odot=0.43$ and $|\cos(\a_0/2)|=0.5$, we need 
$y_1^2\sim 0.5$. 

The important point is that the condition $2\e_e-\e_\tau>0$ is necessary. 
The $\tau$-Yukawa coupling gives $\e_e=0$ and 
$\e_\tau>0$, so that this effect rotates the angle 
into the dark side. To compensate the $\tau$-Yukawa contribution 
and rotates it into the normal side, we need the large neutrino-Yukawa 
couplings and they satisfy 
\begin{eqnarray}
2y_1^2>y_2^2+y_3^2+y_\tau^2\;.
\end{eqnarray}
This is satisfied only when  the neutrino-Yukawa coupling matrix 
has the inverse hierarchical structure. That is, we have to 
assign  
\begin{eqnarray}
y_1&=&m_{D3}\frac{\sqrt{2}}{v\sin \b}\;,\nonumber\\
y_2&=&m_{D2}\frac{\sqrt{2}}{v\sin \b}\;,\nonumber\\ 
y_3&=&m_{D1}\frac{\sqrt{2}}{v\sin \b}\;,
\end{eqnarray}
which fixes the form of $P_{ex}$ in Eq.(14).

\item[(5)] The form of $V_R$

In the diagonal basis of $M=D_R$, we have
\begin{eqnarray}
O_B D_\nu O_B^T =m_D^T D_R^{-1} m_D\;.
\end{eqnarray}
By substituting Eq.(13) and after some computations, we have 
\begin{eqnarray}
M_R^{-1} \equiv (V_R^* D_R^{-1} V_R^\dagger)=
D_D^{-1}(P_{ex}O_B) D_\nu (P_{ex}O_B)^T D_D^{-1}\;.
\end{eqnarray}
We observe that $V_R$ and $M_i$ are determined by 
complex neutrino masses $m_i$ and real Dirac masses $m_{Di}$, 
{\em i.e.}, 6 real positive masses and two Majorana phases. In other 
words, all CP violation phases in this model are related to 
two Majorana phases. 

Now we diagonalize $M_R^{-1}$ in Eq.(25) under the condition 
in Eqs.(8) and (9). In addition, we assume that
\begin{eqnarray} 
m_1\simeq |m_2|\sim |m_3|\;,
\end{eqnarray}
because $m_i$ should be much larger than 
$\sqrt \Delta m_{\rm atm}^2$ 
for the renormalization group to be effective. 
We define
\begin{eqnarray}
a&=&m_1+m_2+2m_3\;\;,\nonumber\\
b&=&m_1+m_2-2m_3\;\;,\nonumber\\
c&=&{\sqrt 2}(m_2-m_1)\;,
\end{eqnarray}
and  
\begin{eqnarray}
\delta_1=\frac{m_{D1}}{m_{D2}}\;,\;
\delta_{2}=\frac{m_{D2}}{m_{D3}}\;,
\end{eqnarray}
which satisfy
\begin{eqnarray}
\delta_1\sim \delta_2\ll 1\;.
\end{eqnarray}

The diagonalization is explicitly given in Appendix, so that 
we give the result. The unitary matrix $V_R$ is given by
\begin{eqnarray}
V_R=\pmatrix{1&-\frac{b}{a}\delta_1
&-\frac{c}{a+b}\delta_1\delta_2\cr
     \left(\frac{b}{a}\right)^*\delta_1&1&- \frac{c}{a+b}\delta_2\cr
     \left(\frac{c}{a}\right)^*\delta_1\delta_2&
     \left(\frac{c}{a+b}\right)^*\delta_2&1 \cr}
    \;,
\end{eqnarray}
where eigenvalues are 
\begin{eqnarray}
\frac{e^{i\phi_1}}{M_1}&=&\frac{a}{4m_{D1}^2}\;,\nonumber\\
\frac{e^{i\phi_2}}{M_2}&=&\frac{1}{4m_{D2}^2}
\frac{4(a+b)m_3}{a}\;,\nonumber\\
\frac{e^{i\phi_3}}{M_3}&=&\frac{1}{4m_{D3}^2}\frac{16m_1m_2}{a+b}\;,
\end{eqnarray}
where phases, $\phi_i$ are defined such that $M_i$ are 
real positive. 
Here we see in general $M_3\gg M_2\gg M_1$ 
because $m_{D3}\gg m_{D2}\gg m_{D1}$, 
except for the very special case $m_1+m_2 =0$.  
It may be interesting to observe that 
\begin{eqnarray}
m_1 m_2 m_3 M_1 M_2 M_3 e^{-i(\phi_1+\phi_2+\phi_3)}
=m_{D1}^2 m_{D2}^2 m_{D3}^2\;.
\end{eqnarray}

As you see in Eqs.(30) and (31), the mixing matrix $V_R$ and also 
the heavy neutrino masses $M_i$ are determined by 
$|m_i|$, two Majorana phases, {\em i.e.}, their relative phases and $m_{Di}$. 
Since the Dirac mass matrix 
is given by $m_D=V_R^\dagger D_D P_{ex}$, phases in it is determined by 
two Majorana phases. 

\end{enumerate}

\section{The numerical analysis}

In the following, we keep only the $y_1$ term 
in the neutrino-Yukawa couplings.   
Also, we take 
\begin{eqnarray}
m_1&=&|m_2|\equiv m\;,\nonumber\\
|m_3|&=&m\pm \frac{\Delta m_{\rm atm}^2}{2m}\;.
\end{eqnarray}
In the following, we take the solar squared mass difference 
given by KamLAND collaboration[12],
$\Delta m_{\rm sol}^2=6.9\times 10^{-5}({\rm eV})^2$ 
and the atmospheric 
squared mass difference  by the SuperKamiokande collaboration[13], 
$\Delta m_{\rm atm}^2=2.5\times 10^{-3}({\rm eV})^2$. The 
atmospheric mixing angle is the maximal as we chose 
the bi-maximal mixing scheme. 

\begin{enumerate}
\item[(1)] The information from the renormalization group analysis

\begin{enumerate}
	\item [(1-a)] The right-handed neutrino mass scale 

If we consider the heavy neutrino threshold correction, $\e_e$ in Eq.(19) 
should be modified by
\begin{eqnarray}
\e_e=\frac{1}{16\pi^2}(Y_\nu^\dagger {\cal L}Y_\nu)_{11}\;,
\end{eqnarray}
where 
\begin{eqnarray}
{\cal L}={\rm diag}(\ln(M_X/M_1),\ln(M_X/M_2),\ln(M_X/M_3))\;.
\end{eqnarray}
We obtain
\begin{eqnarray}
\e_e&=&\frac1{8\pi^2}\frac{m_{D3}^2}{(v\sin \b)^2}
 \ln\left(\frac{M_X}{M_3} \right)\;. 
\end{eqnarray}
For $\epsilon_{\tau}$, $y_3^2-y_2^2$ is negligible because of Eq.(8).
By substituting $\epsilon_e$ and $\e_\tau$ in Eq.(19) into Eq.(21), 
we have
\begin{eqnarray}
\frac{mM_3}{|\cos(\a_0/2)|}\ln\left(\frac{M_X}{M_3}\right)
=\frac{4\pi^2 \cos2\theta_\odot v^2\Delta m_\odot^2 \sin^2 \b 
}{m^2 \cos^2(\a_0/2)}+
\frac{m_\tau^2 \tan^2 \b}2 \ln\left(\frac{M_X}{m_Z} \right)\;,
\end{eqnarray}
where we used $m_{D3}^2\simeq mM_3/|\cos(\a_0/2)|$. 

\begin{figure}
\includegraphics{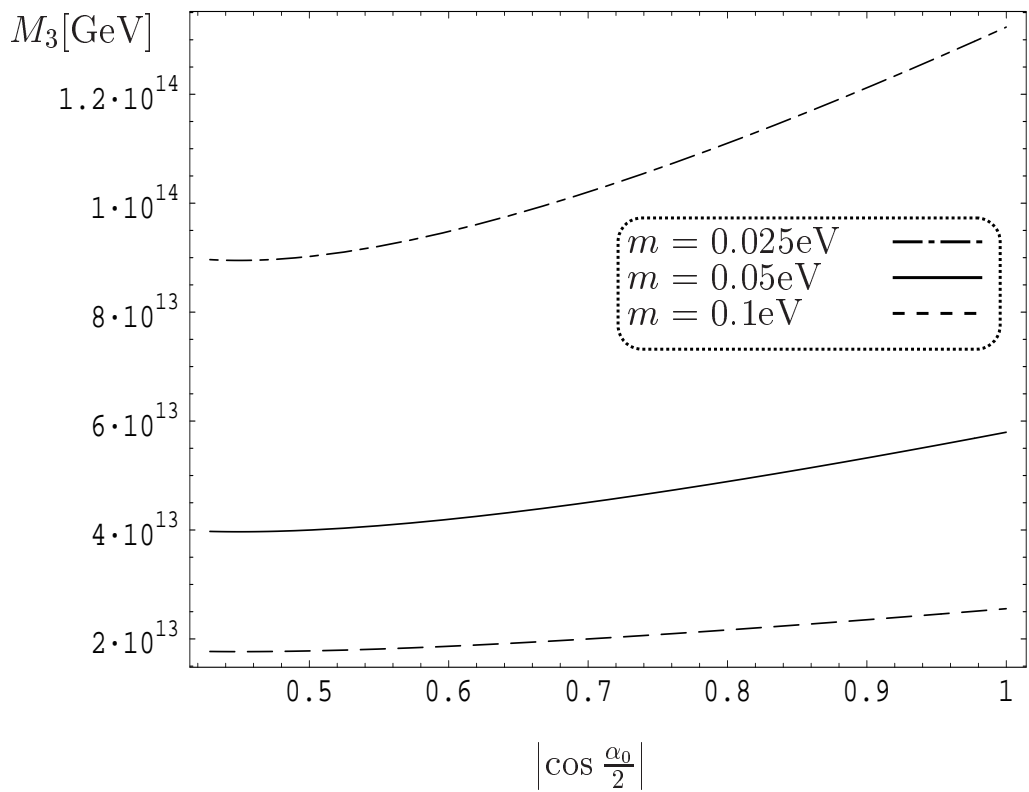}
\caption{
The relation between $M_3$ and $|\cos (\alpha_0/2)|$
for $m=0.025$eV, 0.05eV, 0.1eV. We take $\tan^2\theta_{\odot}=0.4$,
$\tan\beta=20$, and $M_X=10^{16}$GeV.
}
\end{figure}

In Fig.1, we show the relation between $M_3$ and $|\cos(\a_0/2)|$ 
for $m=$0.025eV, 0.05eV, 0.1eV and $\tan^2\theta_\odot=0.4$ with 
$\tan \b=20$, $M_X=10^{16}$GeV. The $M_3$ behaves almost 
independently of $|\cos(\a_0/2)|$ for $m=0.05$eV. 

In particular, 
when 
$|\cos(\a_0/2)|=0.50$, 
$\cos 2\theta_\odot=0.43$ 
and 
$m=0.05$eV, we get
\begin{eqnarray}
M_3 \sim 4 \times 10^{13}{\rm GeV}\;,
\end{eqnarray}
which implies 
\begin{eqnarray}
m_{D3}=\sqrt{\frac{m M_3}{|\cos(\a_0/2)|}}\sim 60{\rm GeV}\;.
\end{eqnarray}
If we take $\tan \b$ larger than 20, $M_3$ becomes larger. 
The other masses $M_i (i=1,2)$ are determined once $m_{Di}$, 
$\a_0$ and $\b_0$ are given.  

\item[(1-b)] The induced $|V_{13}|$

The $|V_{13}|$ and the Dirac CP phase are induced. 
We show only $|V_{13}|$ which is[6] 
\begin{eqnarray}
|V_{13}|=0.010\left(\frac{|\e_\tau|}{9.2\times 10^{-3}} \right)
\left(\frac{m_1m_3}{(0.05)^2{\rm eV}^2}\right) 
\left(\frac{2.5\times 10^{-3}{\rm eV}^2}{\Delta m^2_{\rm{atm}}}\right)
\left(\frac{\sin(\alpha_0/2)}{0.87}\right)
\;,
\end{eqnarray}
where the value $\e_\tau=9.2\times 10^{-3}$ is the one for 
$\tan \b=20$. 
Thus, the model generally predicts the value of $|V_{13}|$ 
which is consistent with the CHOOZ's bound[14] and 
may be detectable in the near future experiments.

\item[(1-c)] The Dirac phase $\delta$

The induced Dirac phase is given by[6]
\begin{eqnarray}
\delta=\frac{\a_0}2-\b_0-\frac{\pi}2+\xi_1+\xi_2\;,
\end{eqnarray}
where $\xi_1={\rm arg}(c-se^{-i\a_0/2})$ and  
$\xi_2={\rm arg}(c+se^{i\a_0/2})$. Here, 
$c=\cos \theta$ and $s=\sin \theta$ and 
\begin{eqnarray}
\sin 2\theta \cos(\a_0/2)=-\cos 2\theta_\odot\;,
\end{eqnarray}
so that 
\begin{eqnarray}
|\cos(\a_0/2)|\ge \cos 2\theta_\odot\;.
\end{eqnarray}

The relation between $\delta + \b_0$ and $|\cos(\a_0/2)|$ is numerically 
plotted in Fig.1 of Ref.6. For 
$|\cos(\a_0/2)|\sim \cos 2\theta_\odot\sim 0.43$, 
$\delta+\b_0$ takes values between $-\pi/2$ and $-3\pi/2$. In 
the discussion of the leptogenesis, we show $\b_0\sim 0$ is favored 
to reproduce the experimental value of the baryon asymmetry. Then, 
for $\beta_0=0$, our model predicts
\begin{eqnarray}
-\frac{1}2\pi>\delta>-\frac32\pi\;.
\end{eqnarray}

\item[(1-d)] The neutrinoless double beta decay

The effective mass[15] of the neutrinoless double beta decay is given by[6]
\begin{eqnarray}
\langle m_{\nu}\rangle
&\simeq& m |\cos(\a_0/2)|\;,
\end{eqnarray}
Since $|\cos (\a_0/2)|\ge \cos 2\theta_\odot$,  $\langle m_{\nu}\rangle>
m\cos 2\theta_\odot\sim 0.43 m$, which may be within the experimental 
sensitivity in the near future. Our expectation is $m\sim 0.05$eV, so 
that it may be around 0.02eV. 
 
\end{enumerate}
\item[(2)] The leptogenesis

The lepton asymmetry parameter, $\e$ is defined by[16]
\begin{eqnarray}
\e&=&\frac{\Gamma(N_1\to \Phi \ell^C)-\Gamma(N_1\to \Phi^\dagger \ell)}
      {\Gamma(N_1\to \Phi \ell^C)+\Gamma(N_1\to \Phi^\dagger \ell)}
      \nonumber\\
   &=&\frac1{4\pi v^2}\frac1{(m_Dm_D^\dagger)_{11}}
   \sum_{j=2,3}{\rm Im}(m_Dm_D^\dagger)^2_{1j}f(M_j^2/M_1^2)\;,   
\end{eqnarray}
where $v\sim 246$GeV and 
\begin{eqnarray}
f(x)\simeq -\frac{3}{2\sqrt{x}}\;.
\end{eqnarray}
The approximate form of $f$ is valid for our case because 
$M_i$  have the hierarchical structure as in Eq.(31). 

By the explicit computation, we find
\begin{eqnarray}
(m_Dm_D^\dagger)_{11}&\simeq& m_{D1}^2\frac{|a|^2+|b|^2+|c|^2}{|a|^2}
\;,\nonumber\\
(m_Dm_D^\dagger)_{12}&\simeq&m_{D1}m_{D2}
\left(\frac{b}{a}+\frac{|c|^2}{a(a+b)^*} \right)e^{i(\phi_1-\phi_2)/2}\;,
\nonumber\\
(m_Dm_D^\dagger)_{13}&\simeq& m_{D1}m_{D3}\left(\frac{c}{a}\right)
e^{i(\phi_1-\phi_3)/2}\;.
\end{eqnarray}
By using Eq.(48),  
$e^{-i\phi_1}=(M_1/4m_{D1}^2)a^*$, 
$e^{-i\phi_2}=(M_2/4m_{D2}^2)(4(a+b)m_3/a)^*$ and
$e^{-i\phi_3}=(M_3/4m_{D3}^2)(16m_1m_2/(a+b))^*$,  
we find 
\begin{eqnarray}
(m_Dm_D^\dagger)_{12}^2\frac{M_1}{M_2}&=& 
 \frac{4m_{D1}^4(a+b)^*m_3^*}{|a|^4}
 \left(b+\frac{|c|^2}{(a+b)^*} \right)\;,\\
(m_Dm_D^\dagger)_{13}^2\frac{M_1}{M_3}&=& 
 \frac{16m_{D1}^4}{|a|^4}\frac{ m_1^*m_2^*a^*c^2}{(a+b)^*}\;,
\end{eqnarray}
Now, we find up to the first order of $\Delta m_{31}^2/m^2$, 
\begin{eqnarray}
\e=\frac{3mM_1}{16\pi v^2}\frac{\cos(\a_0/2)}{R_+}
\left(\frac{\Delta m_{31}^2}{m^2}\right)\sin(\frac{\a_0}2-\beta_0)\;,
\end{eqnarray}
where
\begin{eqnarray}
R_+=\sqrt{1+\cos^2\frac{\a_0}2 
+2\cos\frac{\a_0}2\cos(\frac{\a_0}2-\b_0)}\;,
\end{eqnarray}
and  $\Delta m_{31}^2=\Delta m_{\rm atm}^2$ 
for $|m_3|>|m_1|$ and $-\Delta m_{\rm atm}^2$ for 
$|m_3|<|m_1|$. 
It may be commented that in the approximation of 
$m_1=|m_2|=|m_3|$, the contributions from Eq.(49) and (50) cancel 
each other, so that $\e$ is suppressed by  
$(\Delta m_{\rm atm}^2/m^2)$. 

\begin{figure}
\begin{center}
\includegraphics{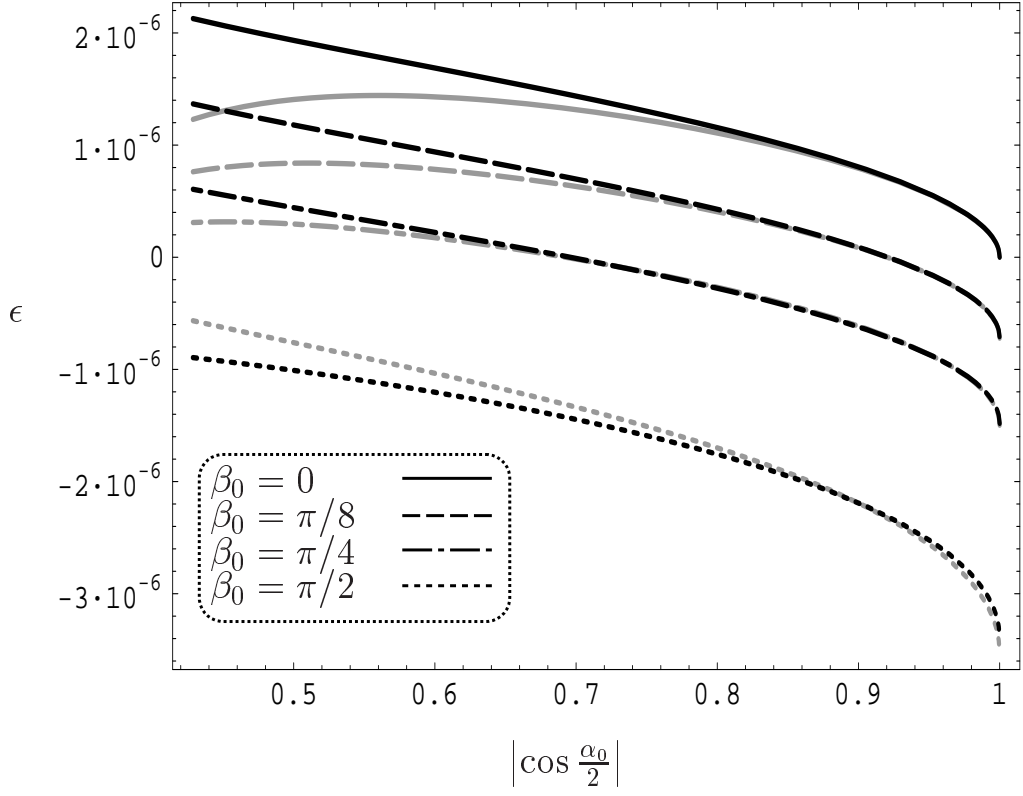}
\end{center}
\caption{The asymmetry parameter $\epsilon$ as a function of 
$|\cos(\alpha_0/2)|$ and $\beta_0$. We take $m=0.05$eV, 
$m_{D1}/m_{D2}=m_{D2}/m_{D3}=1/5$, and $\tan\beta=20$.
The black lines show the exact computations and
the gray lines show the numerical values obtained by
the approximate formula in Eq.~(51).
}
\end{figure}

In Fig.2, the asymmetry parameter $\e$ is plotted as a function of 
$|\cos(\a_0/2)|$ and $\b_0$ with $m=0.05$eV, $m_{D1}/m_{D2}=
m_{D2}/m_{D3}=1/5$, and $\tan \b=20$. The larger $\epsilon$ is obtained for 
smaller $|\cos(\a_0/2)|>\cos 2\theta_{\odot}$ and also $\b_0$. 
The black lines show 
the numerical computation without any approximation and the 
gray lines are obtained by using our approximate formula 
in Eq.(50). For $|\cos(\a_0/2)|<0.6$, there are some difference between 
the exact computations and Eq.(51).  
We can obtain  
$\e\sim 10^{-6}$ for $\b_0\sim 0$ and 
$\cos 2\theta_\odot <|\cos(\a_0/2)|<0.8$. 

The baryon asymmetry parameter is given by[17]
\begin{eqnarray}
\eta_{B0}\simeq -10^{-2}\e\kappa_0\;,
\end{eqnarray}
where for small $m$, $\kappa_0\simeq 1/(2\sqrt{K^2+9})$ with 
$K\sim 170(m/{\rm eV})$. With $m=0.05$eV, we find $\kappa_0\sim 
6\times 10^{-2}$, so that we obtain
\begin{eqnarray}
\eta_{B0}\sim 6\times 10^{-10}\;,
\end{eqnarray}
which agrees with the experimental value[18]. 

Of course, the value of the asymmetry parameter depends 
on $M_1$ linearly, which we derived by assuming 
$m_{D1}/m_{D2}=m_{D2}/m_{D3}=1/5$. If the hierarchy of 
the eigenvalues of the Dirac mass matrix is larger than 
what we used, we find the smaller $M_1$ which results in 
the smaller asymmetry parameter. In other words, we can 
explore how hierarchical the Dirac masses 
are from the asymmetry parameter $\epsilon$.

\item[(3)] The LFV processes

In this model, the LFV processes take place through the 
slepton mixing, which is absent at the GUT scale. 
However, the slepton mixing is induced 
by the renormalization group effects
at the scale $M_R$ where the right-handed 
Majorana neutrinos are decoupled.
In the leading log approximation, the off-diagonal terms of the 
slepton mass matrix is given by[18]
\begin{eqnarray}
(m_{\tilde{L}}^2)_{ij}\simeq \frac{6m_0^2+2|A_0|^2}{16\pi^2}
(Y_\nu^\dagger {\cal L}Y_\nu)_{ij}
\;,
\end{eqnarray}
for $i\neq j$, where ${\cal L}$ is given in Eq.(35). 
The off-diagonal elements contributes to lepton flavor violating
processes such as $\mu\to e\gamma$. The decay width of $l_i\to l_j\gamma$
process is approximately given by[18]
\begin{eqnarray}
\Gamma(l_i\to l_j\gamma)&\sim&
\frac{\alpha^3m_{l_i}^5}{192\pi^3}
\frac{|(m_{\tilde{L}}^2)_{ij}|^2}{m_S^8}\tan^2\beta\nonumber\\
&=&\frac{\alpha^3m_{l_i}^5}{12(4\pi)^5}\frac{(6m_0^2+2A_0^2)^2}{m_S^8}
\frac{\tan^2\beta}{(v\sin\beta)^4}|(m_D^{\dagger}{\cal L}m_D)_{ij}|^2\;,
\end{eqnarray}
where $m_S$ represents typical mass of supersymmetric particles.

It is convenient to separate $m_D^{\dagger}{\cal L}m_D$ into 
two parts as 
\begin{eqnarray}
m_D^\dagger {\cal L}m_D =  <{\cal L}> m_D^\dagger m_D
+m_D^\dagger ({\cal L}-<{\cal L}>)m_D \;, 
\end{eqnarray}
where 
\begin{eqnarray}
<{\cal L}>&=&\ln(M_X/M_2){\rm diag}(1,1,1)\;,\nonumber\\
{\cal L}-<{\cal L}>&=&{\rm diag}
(\ln(M_2/M_1),0,\ln(M_2/M_3))\;.
\end{eqnarray}
The mass $M_{2}$ is considered as a kind of average of $M_{i}$. 
Usually, the 1st term of the right-hand side of Eq.(57) is considered. 
However, our model gives $(Y_\nu^\dagger Y_\nu)_{ij}=0\; (i\neq j)$ 
because of the assumption (A.3), and thus
the  LFV processes occur only through the 2nd term. 

In this model, the off-diagonal elements of 
$m_D^{\dagger}\mathcal{L}m_D$ become
\begin{eqnarray}
\left|(m_D^{\dagger}m_D)_{12}\right|&=&
m_{D2}^2\left|\frac{c}{a+b}\right|\ln\frac{M_2}{M_3}\;,\\
\left|(m_D^{\dagger}m_D)_{23}\right|&=&
m_{D1}^2\left|\frac{b}{a}\right|\ln\frac{M_1}{M_2}\;,\\
\left|(m_D^{\dagger}m_D)_{13}\right|&=&
m_{D1}^2\left|\frac{c}{a}\right|\ln\frac{M_1}{M_3}\;.
\end{eqnarray}
Thus, 
$\tau\to \mu \gamma$ and $\tau\to e\gamma$ processes 
are suppressed by 
factor $(m_{D1}/m_{D2})^2$ in comparison with 
the $\mu\to e\gamma$. 
While $\mu\to e\gamma$ is independent of $\beta_0$,
$\tau\to \mu\gamma$ and $\tau\to e\gamma$ depend
on $\beta_0$ as well as $\a_0$. 
We find the branching ratios of $\ell_i\to \ell_j\gamma$ become larger
as $\cos(\alpha_0/2)$ become smaller.
In the limit of $\cos(\alpha_0/2)=1$, $\mu\to e\gamma$ and $\tau\to e\gamma$
do not occur, but $\tau\to \mu\gamma$ can occur if 
$\beta_0$ is not 0 or $\pi$.

For almost all values of $\cos(\alpha_0/2)$ and $\beta_0$, 
we obtain  
\begin{eqnarray}
\rm{Br}(\tau\to \mu \gamma)<
\rm{Br}(\tau\to \mu e)<
\rm{Br}(\mu\to e \gamma)
\;.
\end{eqnarray}
However, the branching ratios of these processes except for 
$\mu\to e\gamma$ are 
too small to be observed in the 
future experiments. Their typical values are 
 $\rm{Br}(\ell_i\to \ell_j\gamma)<10^{-12}$.
\end{enumerate}

\section{Comments and discussions}

We constructed a model that all CP violation phases 
at the high energy scale and the low energy scale are controlled by 
the two Majorana phases which appear as the relative 
phases of neutrino masses. This strong restriction of the 
model is due to the assumption that $Y_\nu^\dagger Y_\nu$ is 
diagonal. This requirement is motivated by the consideration 
that the large diagonal elements are needed to reconcile the 
maximal solar mixing angle at GUT scale and the observed one 
at the low energy scale. We found that the induced $|V_{13}|$ is 
the measurable size in the near future experiments, 
the induced Dirac CP phase is likely in between $-\pi/2$ and 
$-3\pi/2$, the asymmetry parameter can be of order $10^{-6}$, 
the LFV processes are suppressed. 
In this paper, we considered $\tan\beta=20$ case. For smaller 
$\tan\beta$, $M_3$ becomes smaller as we see in Eq.(37).
Then, $M_1$ becomes smaller too.
Since $\epsilon$ is proportional to $M_1$ in this model, 
$\epsilon$ becomes smaller, so that it would become hard to 
explain the baryon number asymmetry in the universe.

Of course, the assumption $Y_\nu^\dagger Y_\nu$ may be too strong. 
The renormalization group argument for the solar mixing parameters 
requires that the $(Y_\nu^\dagger Y_\nu)_{11}$ elements must be much 
larger than the other elements. Therefore, a general form 
would be 
\begin{eqnarray}
Y_\nu^\dagger Y_\nu=y_1^2\pmatrix{1 &*&*\cr *&*&*\cr *&*&*\cr}
\nonumber\\
\end{eqnarray}
where $y_1^2$ is of order 0.5 and elements shown by * are 
of order $10^{-4}$. Although these elements are small, 
they will contribute to the LFV processes. However, their 
sizes are not controlled by the model and 
we lost the predictions for them. 

We can construct a similar model to the present model, by 
assuming the eigenvalues of the 
Dirac mass matrix 
are quasi-degenerate. This analysis will be reported soon.

\vskip 5mm
{\Huge Acknowledgment} 
This work is supported in part by 
the Japanese Grant-in-Aid for Scientific Research of
Ministry of Education, Science, Sports and Culture, 
No.12047218 and No.15540276.

The work of T.S. was supported in part by Research Fellowship of the
Japan Society for the Promotion of Science (JSPS) for
Young Scientists (No.15-03927).

\newpage
\setcounter{section}{0}
\renewcommand{\thesection}{\Alph{section}}
\renewcommand{\theequation}{\thesection .\arabic{equation}}
\newcommand{\apsc}[1]{\stepcounter{section}\noindent
\setcounter{equation}{0}{\Large{\bf{Appendix\,\thesection:\,{#1}}}}}

\apsc{Detailed derivations}

We parametrize $D_D^{-1}$ as
\begin{eqnarray}
D_D^{-1}=\frac1{m_{D1}}\pmatrix{Q&0\cr 0&\delta_1\delta_2\cr}\;,
\end{eqnarray}
where $\delta_i$ are defined in Eq.(28) and 
\begin{eqnarray}
Q=\pmatrix{1&0\cr 0&\delta_1 \cr}\;\;.
\end{eqnarray}
We find by using $a$, $b$ and $c$ defined in Eq.(27),
\begin{eqnarray}
M_R^{-1}\simeq \frac{1}{4m_{D1}^2}
\pmatrix{QXQ&
\delta_{1}\delta_2 QY\cr \delta_{1} \delta_{2}Y^TQ&
\delta_{1}^2\delta_2^2 (a+b) \cr}\;,
\end{eqnarray}
where
\begin{eqnarray}
X=\pmatrix{a&b\cr b&a\cr}\;,\;\; Y=c\pmatrix{1\cr 1\cr}
\;\;.
\end{eqnarray}

We block diagonalize the matrix $M_R^{-1}$, 
by the seesaw calculation with respect to $\delta_1\delta_2\ll 1$. 
We remind that $a$, $b$, $c$ are 
quantities of the same order because 
$|m_1|\simeq |m_2|\sim |m_3|$. 
We find with
\begin{eqnarray}
V_1=\pmatrix{1&-\delta_{1} \delta_{2}Q^{-1}X^{-1}Y\cr 
\delta_{1} \delta_{2} (Q^{-1}X^{-1}Y)^\dagger&1\cr}\;,
\end{eqnarray}
\begin{eqnarray}
V_1^T M_R^{-1} V_1 \simeq \frac{1}{4m_{D1}^2}
\pmatrix{ QXQ&0\cr
0& \delta_1^2 \delta_{2}^2(a+b-Y^TX^{-1}Y)\cr}\;
\end{eqnarray}

Next, we diagonalize  $QXQ$ in the first order of the small 
quantity $\delta_{1}$ by applying 
\begin{eqnarray}
V_2=\pmatrix{1&-\delta_{1}(b/a)&0\cr
   \delta_{1}(b/a)^*&1 &0\cr 
   0&0&1\cr}\;\;,
\end{eqnarray}
and we find that $M_R^{-1}$ is diagonalized as
\begin{eqnarray}
(V_1V_2)^T M_R^{-1} (V_1V_2)
={\rm diag}\left(\frac{e^{i\phi_1}}{M_1},\frac{e^{i\phi_2}}{M_2},
\frac{e^{i\phi_3}}{M_3} \right)\;,
\end{eqnarray}
where $M_i$ and phases are given in Eq.(31).

\end{document}